\documentclass[onecolumn,aps,superscriptaddress]{revtex4}

\usepackage[english]{babel}
\usepackage[pdftex]{graphicx,graphics}
\usepackage{amssymb,amsfonts,amsmath}
\usepackage{multirow}
\usepackage{color}

\begin{document}

\title{Time as a limited resource: Communication Strategy in Mobile Phone Networks}

\author{Giovanna Miritello}
\affiliation{Departamento de Matem\'aticas \& GISC, Universidad Carlos III de Madrid, 28911 Legan\'es, Spain}
\affiliation{Telef\'onica Research, 28050 Madrid, Spain}

\author{Esteban Moro}
\affiliation{Departamento de Matem\'aticas \& GISC, Universidad Carlos III de Madrid, 28911 Legan\'es, Spain}
\affiliation{Instituto de Ingenier\'{\i}a del Conocimiento, Universidad Aut\'oma de Madrid, 28049 Madrid, Spain}

\author{Rub\'en Lara}
\affiliation{Telef\'onica Research, 28050 Madrid, Spain}

\author{Roc\'io Mart\'inez-L\'opez}
\affiliation{Telef\'onica Research, 28050 Madrid, Spain}

\author{John Belchamber}
\affiliation{Telef\'onica Research, 28050 Madrid, Spain}

\author{Sam G. B. Roberts}
\affiliation{Department of Psychology, University of Chester,  Cheyney Road, Chester CH1 4BJ}
\author{Robin I.M. Dunbar}
\affiliation{Department of Experimental Psychology, University of Oxford, South Parks Road, Oxford OX1 3UD UK}

\date{\today}

\begin{abstract}
We used a large database of 9 billion calls from 20 million mobile users to examine the relationships between aggregated time spent on the phone, personal network size, tie strength and the way in which users distributed their limited time across their network (disparity). Compared to those with smaller networks, those with large networks did not devote proportionally more time to communication and had on average weaker ties (as measured by time spent communicating). 
Further, there were not substantially different levels of disparity between individuals, in that mobile users tend to distribute their time very unevenly across their network, with a large proportion of calls going to a small number of individuals. Together, these results suggest that there are time constraints which limit tie strength in large personal networks, and that even high levels of mobile communication do not fundamentally alter the disparity of time allocation across networks.
\end{abstract}


\maketitle

\section{Introduction}
\label{Intro}
During the last two decades the structural and the dynamic properties of social networks have been subject of intensive study (\citealt{Watts1}). The structure of social networks is important not only from the perspective of the single user, but also from that of society as a whole, as it can influence various dynamic processes of human interaction, communication, spreading of information and disease transmission (\citealt{Christakis, Onnela1, Watts1}). Traditionally, these communication networks have been studied on a relatively small scale, using questionnaire or interview methods to gather data on how communication patterns are related to other characteristics such as social support (\citealt{Wellman3})  or the emotional intensity of the tie between two individuals (\citealt{Roberts3}). However, people's recollection of specific communication events is often imperfect (\citealt{Bernard1}) and the extent to which studies on specific, limited samples can be generalized to wider populations and countries is unclear (\citealt{Henrich, Wellman1}).
With the rise of electronically-mediated communication, it is now becoming possible to study communication patterns in networks on a scale, and at a level of detail, not possible using traditional questionnaire or survey methods (\citealt{Bohannon, Lazer, Watts2}). Specifically in terms of mobile phone communication, access to data on this scale has led to advances in our understanding about the structure of mobile phone networks, network dynamics, factors influencing information transmission in the networks and reciprocity of communication (\citealt{Miritello, Onnela2, Palla}). 

One key variable that characterizes the structural topology of such networks is the social connectivity or degree of a node.  It measures the number of people with whom an individual interacts and can also be interpreted as a measure of social integration (\citealt{Marsden}) or activity (\citealt{Wasserman}).
In general, the degree distributions are skewed with a long tail, indicating that most users have only a few connections while a small minority have hundreds of connections (\citealt{Newman}).  Social connectivity, however, varies depending on the nature of ties. In fact, within the same network, not all the connections have the same importance/role. 
For this reason, in several networks and in social networks in particular, each tie has a given strength that quantifies the attention or the flow of information through that connection. This is a reflection of real life, where people maintain a large number of relationships with a different strength or importance: family, friends, work colleagues, acquaintances (\citealt{Granovetter1, Roberts2, Wellman4}). Exploring the strength of the ties in social networks can help in the understanding of the structure of the network but also of the dynamics of many phenomena that involve human behaviour such as the formation of communities, the spreading of information and social influence (\citealt{Hill1,Onnela1,Watts1}).

However, what has not been comprehensively explored in these studies is the way in which the size of an individualÕs personal network - their set of ties to family and friends (\citealt{Wellman2}) - affects the way in which they distribute their time across that network. Time is an inelastic resource, and people only have a limited amount of time in each day to devote to social interaction (\citealt{Nie, Roberts2}). Further, the emotional intensity of a tie is strongly related to the frequency of communication between the two individuals (\citealt{Hill2, Roberts3, Wellman4, Saramaki}) and frequent communication is necessary to prevent a tie from decaying in emotional intensity over time (\citealt{Cummings, Lee, Oswald, Roberts4}). Since communication always takes time, the limited amount of time available for  communication acts as a constraint on the number of ties that can be maintained at each level of emotional intensity (\citealt*{Roberts2, Roberts1}).  
However, mobile communication offers a greater ease of communication than face-to-face interaction, and the total volume of communication by mobile phone has increased markedly, even in the last decade. In the UK, between 2004 and 2009 the number of mobile voice call minutes rose from 64 billion to 118 billion, and this was accompanied by only a marginal decline in fixed voice line minutes from 163 billion to 113 billion (\citealt{Ofcom}). The number of text messages sent rose from 26.9 billion to 103.9 billion over the same period, which means that by 2009 each person in the UK was sending an average of over four text messages a day.

The key question we address in the present study is how people distribute their activity in communication across their personal network. How does the ease with which people can communicate over mobile phones affect the way in which they distribute their limited amount of time across their social network? 
We address this question by exploring the relationship between the time spent on voice calls over a given period (the "intensity of use") and the way in which these calls are distributed across the social network. In general, people do not distribute their time evenly  across their social network, but tend to focus the majority of their time on a relatively small number of ties. Thus, for example, although people may have hundreds of ÔfriendsÕ on Facebook, they are only in direct contact with a small proportion of these friends over a given time period (\citealt{Marlow}). Similarly, people tend to be in much more frequent contact with those with whom they are emotionally close (\citealt{Hill2, Roberts3}), and mobile users tend focus a large proportion of their communication on a small number of people (\citealt{Reid}).
However, this latter study was based on self-report data and there has been a large increase in the volume of communication by mobile phone since this study was conducted. 

This paper examines these patterns using a much larger, more representative sample reflecting current mobile phone usage. Because the analysis is based on detailed records of actual mobile communication, this paper also avoids the problems of accuracy and reliability of self-report data (\citealt{Bernard1}).
Further, the ease of communication over mobile phones may offer an opportunity to diverge from this typical pattern. 
If there is a greater volume of communication over mobile phones, are people using this greater volume to communicate with their strong ties, with whom they tend to have frequent contact anyway, or to build up their ÔweakÕ ties? Strong, emotionally intense ties are crucial in providing emotional and material support (\citealt{Plickert, Wellman2}) and for physical health and wellbeing both in humans (\citealt*{Holt}) and in non-human primates (\citealt{Silk,Wittig}). However, weak ties provide access to a greater variety of information than strong ties (\citealt{Eagle2, Granovetter1}) and thus potentially people may use the ease and convenience of mobile communication to focus on strengthening their weak ties, rather than communicating even more intensively with their strong ties. 
There is an active debate as to the extent to which a greater ease of communication (whether over mobile phones or via the internet) fundamentally alters patterns of human sociality. Some authors have argued that a greater ease of communication may allow users to strengthen their weak ties (\citealt{Donath}), whilst other have argued that electronic communication tends to supplement, rather than reshape, existing patterns of sociality (\citealt{Dunbar2}). 

In terms of mobile phone use, broadly there are three possibilities for how intensity of use may affect distribution of time across the network. 
First, if people are focusing on building up their Ôweak tiesÕ, then we may expect them to use the ease of mobile communication to broaden their social network and thus to spread their effort relatively evenly across their network. If this is the case, we would expect that people with a high intensity of use to show a more homogenous pattern of communication across their network, rather than focusing most time on a few close ties. Second, if people  are focusing on building up their strong ties, we may expect them to use their mobiles to deepen the ties in their social network and thus to spread their effort relatively unevenly across their network. If this is the case, we would expect even those with a high intensity of mobile use to show a very uneven spread of effort across their network. 
Third, if people simply use their mobiles to facilitate how they would interact via other communication modes, we may expect the intensity of mobile use not to be associated with changes in the spread of effort across the social network ('no change' model).  
  
In this study, we use a very large dataset of 9 billion calls from 20 million mobile users from a single country over 11 months to examine to examine how individuals' communication strategies (the way in which individuals distribute their limited time across their social network) are related to the size of their personal network and the intensity of mobile use. 
We use the time mobile users spend calling others as a rough estimate of the strength of the tie to that individual, which is strictly related to the frequency with which two individuals communicate. 
Specifically, the frequency with which two individuals communicate has been found to be related with the emotional intensity of a relationship (\citealt{Hill2, Roberts3}). Further, a comprehensive analysis on the potential predictors of tie strength in Facebook users revealed that the intensity of the communication (volume of communication between two individuals) and the frequency of communication (days since last communication) were two of the best predictive variables (\citealt{Gilbert, Bakshy}). As time is inevitably limited, is seems reasonable to suppose that the way in which people differentially invest their time across their connections reflects, to some extent, the value they place on those relationships. 

Previous research using mobile phone databases has tended to either focus on processes at the whole network level (\citealt{Onnela2,Palla}) or at the level of dyadic ties (\citealt{Hidalgo,Reader}). However, there is an increasing interest in using the digital trace of electronic communication (\citealt{Lazer}) to study personal (or egocentric) networks, which represent an important intermediate level linking dyadic ties and the whole network (\citealt{Wellman2}). Studying mobile networks at the level of personal networks allows important questions to be addressed concerning the factors that shape communication patterns of individuals, and how these communication patterns are influenced by the size of the personal network, and the intensity of mobile phone use. Whilst these personal networks have been studied for online communication (\citealt{Arnaboldi,Gilbert,Passarella}), there have been few studies examining personal networks based on mobile records, and those studies that do exist have tended to study only very small parts of the personal network (the three "best friends", \citealt{Palchykov}), or been based on a limited sample (94 participants, \citealt{Eagle1}).

Thus this work extends previous studies in this area in two key ways. First, we carry out a study of personal networks on a very large scale to examine the relationship between the size of an individual's personal network and the constraints on communication time, an aspect that is crucial to understanding the dynamics of social relationships (\citealt*{Sutcliffe}), but has previously only been investigated using questionnaire data (\citealt{Roberts1,Roberts4}). 
Second, we explore the relationship between intensity of mobile use and the evenness with which people distribute their time across their social network, to address fundamental questions about whether a higher volume of communication affects the typical uneven distribution of communication seen in personal networks, and thus whether technology use more broadly may affect underlying patterns of human sociality (\citealt{Dunbar2,Donath}). Further, all the results are tested against null models to distinguish whether there is a genuine communication strategy within personal networks, or whether the same results can be obtained with a network in which the time each user dedicates to any of their connections does not correspond to the real one.

\section{Dataset}
\label{Data}

We analyze the social network formed by the mobile Call Detail Records (CDRs) from a single European operator over a period of 11 months. In order to maintain privacy, the records were anonymous so we did not have access to the names of the users or to their phone numbers and only an identification number was provided to characterize each user.
The database consisted of over $20$ million users and $9$ billion calls between users and contained several fields about each call. However, for the purposes of this paper we only focus on the source and the destination identification numbers and the duration of calls. We filtered out all the calls involving other operators, keeping only those events in which the calling and the receiving number belong to the operator under consideration and we consider only voice communication events, thus not including short text and multimedia messages (SMS/MMS). 
We focused on voice communication in this study because we were particularly interested in the time dimension - the amount of time people dedicate to each of their social ties - and voice calls have a duration, whereas text messages do not.
Further, mobile voice communication has been shown to accurately predict self-reported friendships (\citealt{Eagle1}). Whilst text messages also have considerable scope for uncovering friendship patterns (\citealt{Reid}), including them in a separate analysis is beyond the scope of this paper. 
Due to the bidirectional nature of phone communication, we consider a call from $i$ to $j$ equivalent to a call from $j$ to $i$. To avoid business numbers and voice operator calls we only consider reciprocated ties (\citealt{Onnela2}), thus one undirected link between node $i$ and node $j$ is established if at least one pair of communication calls ($i \rightarrow j$ and $j \rightarrow i$) is observed during the whole time period.

\section{The boundaries of human communication}
\label{Diversity}
From the data described above we study the weighted social network, where the weight (or intensity) $w_{ij}$ of an edge connecting user $i$ and user $j$ is defined as the aggregated time the two users spend talking to one another.
Usually, in communication networks the tie weight $i \leftrightarrow j$  is taken as the total number of calls or, analogously, as the aggregated duration of calls between $i$ and $j$ during the period under investigation. 
These two quantities are strongly related and  previous studies have shown that they give an equivalent quantification of the tie weight (\citealt{Onnela2}).  
For a given user $i$ the social connectivity $k_i$ is defined as $k_i=\sum_j a_{ij}$, where $a_{ij}$ is equal to 1 if a connection between $i$ and $j$ exists and 0 otherwise.  The sum of $w_{ij}$ over all his neighbors $s_i=\sum_j a_{ij}~w_{ij}$ defines the strength or intensity $s_i$ of the user, which is a measure of his strength in terms of the total weight of his connections. 
Both the social connectivity and the strength $s_i$ are very heterogeneous with a highly skewed distribution (\citealt{Onnela2}). In our database, we observe a mean social connectivity of 85.2, with a median of 62 and maximum value of around 500.
For the node strength $s_i$ instead, we found that, although the mean of this distribution is around 1.5 hours in the whole period, the maximum value is about  6  hours per day. This means that while the time that the larger part of the population spends on the phone per day is of the order of seconds or minutes, there is a small minority who phone more than 1 hour per day.
Not only the aggregated $s_i$, but also $w_{ij}$ show a long-tailed distribution across the whole population (\citealt{Onnela2}).  As in most studies on communication networks, this result refers to the population as a whole. However,  due to the limited amount of time people can devote to social relationships, it is possible that each user follows a different strategy to allocate his time across their contacts, according to the total size of his personal network or the total time he spends in phone communication. In this case, the observed heterogeneity of $w_{ij}$ could be due to the existence of users with different communication strategies. 

\begin{figure}[t]
\centering
\includegraphics[width=1\textwidth]{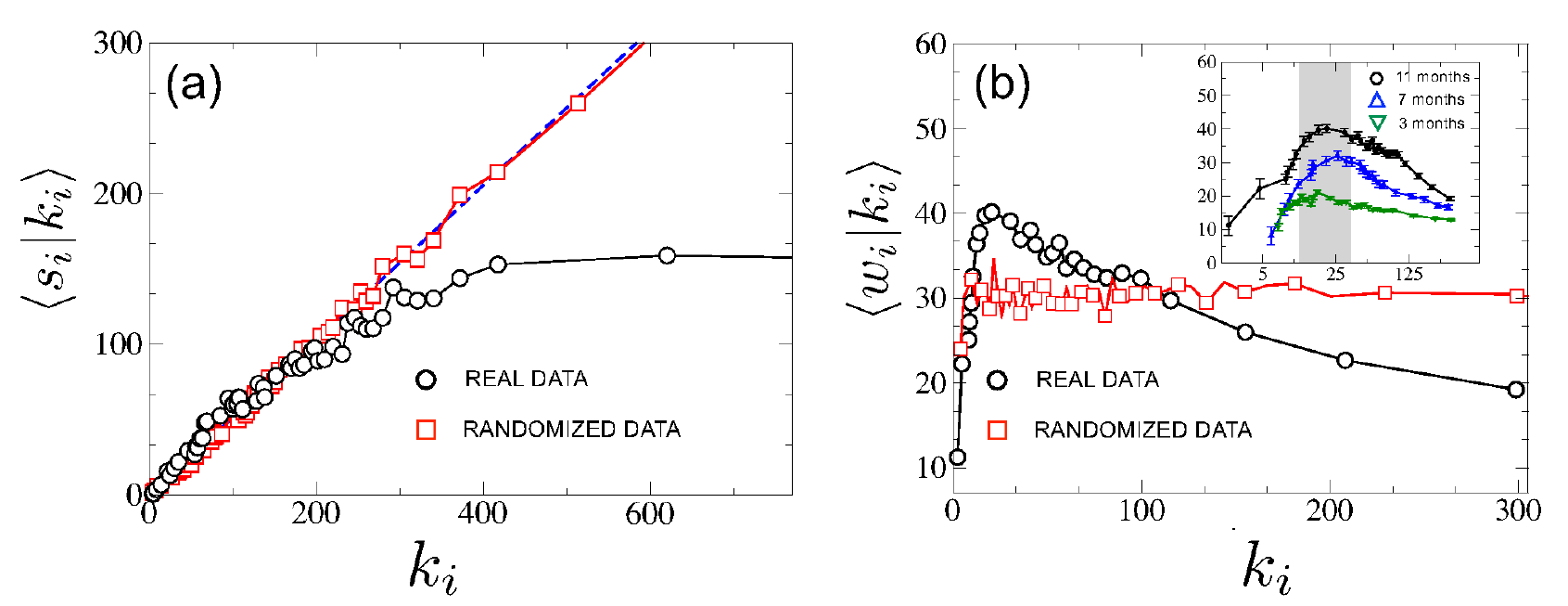}
\caption{(color online)
(a) Average strength $s_i$ of a node (measured as the aggregated duration, in hours, of phone calls) for a given social connectivity as a function of the social connectivity $k_i$.  The open circles correspond to the real data and the open squares to the randomised network. 
For real data the strength of nodes saturates for relatively large values of connectivity, while in the randomised case the data can be fitted by the uncorrelated approximation $s_{i} = \langle w \rangle ~ k{i} $, corresponding to the blue dashed line. (b) Average weight of the ties (in hours) as a function of the social connectivity. For the real data the average weight of each tie gradually increases with the connectivity until it reach a maximum, which is estimated to be around 20 connections. No significant dependence is observed in the randomised case. In the inset we compare the results obtained in the observation time period of 11 months (open black circles) with the ones obtained for a period of 3 and 7 months (green down-side and blue up-side triangles respectively).}
\label{fig1}
\end{figure}

To address this, we first analyze how the strength of a node varies as a function of its social connectivity and compare the results with the randomised network.
In the latter, the weight of each tie is replaced by a randomly selected tie weight from the whole network. Note that in the randomised network the overall social connectivity of each user (and thus the network topology) is preserved, while the amount of time each user dedicates to all his connections does not now correspond to the actual value. 
As shown in Fig.\ref{fig1} (a), we observe that the strength $s_{i}$ of a node increases with the total number of its connections $k_{i}$: thus people with many contacts invest much more time in communication than people with few of them. This result is in line with previous studies (\citealt{Barrat}), where the authors find that the average strength $s(k)$ of nodes with degree $k$ increases with $k$ as $s(k)\sim k^{\beta}$ for both scientific collaboration and the air-transportation networks.
Specifically, we observe that the average strength of a node for a given $k_{i}$ increases with $k_i$, then saturates for large values of  $k_i$, which suggests the existence of a limit in the user's ability to communicate. The latter reflects the fact that time is finite, thus if users add people to their network, the time they invest to communicate with them does not necessarily increase in a proportional fashion, which Fig. 1 (a) demonstrates applies to people with very large social networks (approximately 300).
In the same figure we also show the results obtained for the randomised network. In this case the average strength of a node with a given connectivity $k_{i}$ is exactly proportional to $k_{i}$, yielding an exponent $\beta = 1$.  This means that given the number of connections of a node, the corresponding average strength is provided as well. In this case $s_i$ is very well fitted by the approximation $s_i=\langle w \rangle~k_i$, which indicates that the tie weights are mostly not correlated to the degree of the node $i$. In fact, in the absence of correlations between the tie weights and the social connectivity, we can approximate $w_{ij}=\langle w \rangle$, where $\langle w \rangle$ is the average tie weight in the whole network.
The deviations from such linear behaviour observed for real data suggest the existence of correlations between $s_i$ and $k_i$.
In particular, the fact that $s_i$ increases sub-linearly for large $k_i$ indicates that, on average,  highly connected people tend to spend less time on the phone that the one that they would spend with a random assignment of weights. 
This behaviour differs from the one observed for the air-transportation network (\citealt{Barrat}), where a super-linear behaviour is observed, meaning the airport traffic (which defines the node strength) grows faster than its degree (number of direct flights). 
While for the air-transportation network the larger is the airport the more traffic it can handle, in the case of mobile-phone network a larger number of contacts does not necessarily imply a larger investment of time (and money) in communication.

The existence of a limit in the communication time is better observed by looking at the average weight of ties of each user, which we refer as $w_i=\sum_{j=1}^{k_i} w_{ij}/k_i=s_i/k_i$, as a function of the social connectivity $k_i$.
This is shown for both the real and the randomised network in Fig.\ref{fig1} (b).
Here it becomes much more clear that, on average, the time dedicated to each connection gradually increases with $k_i$: the more relationships people have, the more time they need to dedicate on average to each of them. However, when the number of connections surpasses a certain threshold, which is around $k_i \in[10,40]$, the user can no longer dedicate he same amount of time to each of them. This is why the average value of $w_{ij}$ reaches a maximum, then starts to decrease with $k_i$.
Interestingly, we find that the position of the peak does not change with the length of the time period. This is shown in the inset of Fig. \ref{fig1} (b), where we compare the results obtained in the time period of 11 months with the observation within 3 and 7 months. This finding demonstrates the robustness of our result and shows that the limit in people's social capacity is an intrinsic constraint and not an effect of the finite time window.
\begin{figure}[t] 
\centering
\includegraphics[width=0.6\textwidth]{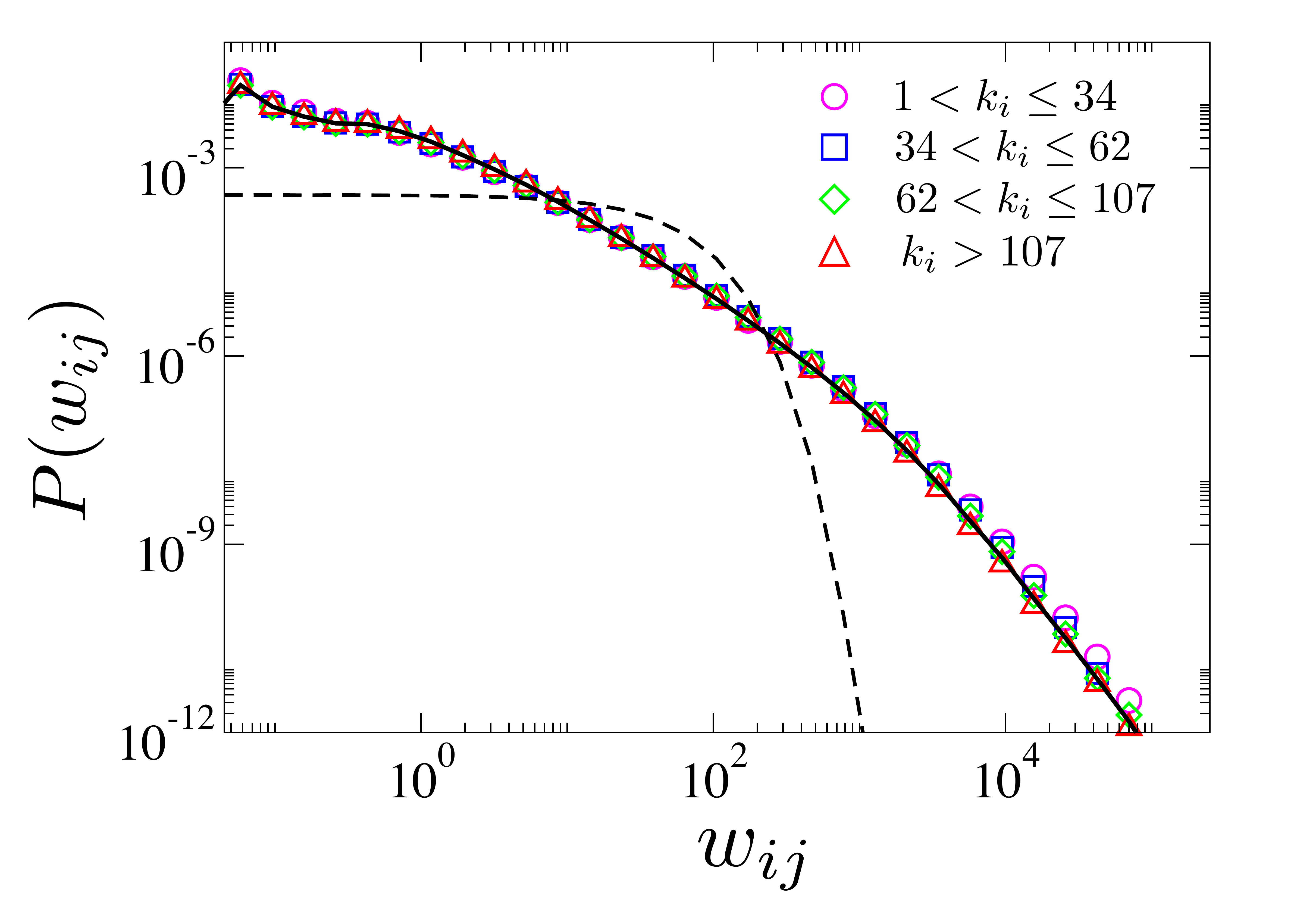}
\caption{(color online) Distribution of the ties weights for nodes with different values of social connectivity that correspond to the quartile of the distribution of $k$: 1$<$ k $\leq$ 34 (magenta circles), 34 $<$ k $\leq$ 62 (blue squares), 62$<$ k $\leq$ 107 (green diamonds) and k $\geq$ 107 (red triangles) compared to the one obtained when the whole population is considered (black solid curve). The distribution does not show a significant dependence on the social connectivity. For comparison, we also show an exponential distribution with the same mean of the real distribution (black dashed curve).}
\label{fig2}
\end{figure}

\section{Time allocation diversity}
The results in Fig.\ref{fig1} indicate that people with larger networks dedicate on average less time to each of their social connections. This raises the question on whether the heterogeneity observed in the distribution of tie weights is due to the communication strategy being dependent on the social connectivity. For example, one might assume that users with a large number of connections contribute to small weights more than users with few connections. However, we found that the distribution of $w_{ij}$ does not show an appreciable dependence on the social connectivity. This result is shown in Fig.\ref{fig2}, where we show the tie weights distribution. Each curve corresponds to the tie weights distribution for nodes belonging to different intervals of social connectivity, chosen accordingly to the quartiles of the whole distribution of $k_i$ (only users with $k_{i}>1$ are considered). 
This finding indicates that, on average, people distribute their time unevenly across their contacts: they dedicate a small amount of time to many people and a large amount of time to a small number of people, independently of the size of their social circle.

To investigate users' diversity in time allocation in more detail, we measured the \emph{disparity} $Y_{i}$ , which constitutes a widely used measure of diversity in the network literature (\citealt{Boccaletti,Barthelemy1,Barthelemy2}) and is given by:
\begin{equation}
Y_{i} = \sum_{j=1}^{k_{i}}\left(\frac{w_{ij}}{s_{i}}\right)^2.
\label{eq-disp}
\end{equation}
The disparity is a measure of local heterogeneity. In the homogeneous case, in which a user maintains a even communication with all his neighbors, $Y_{i}\simeq1/k_{i}$, since $w_{ij}=s_{i}/k_{i}$. In contrast, in presence of perfect heterogeneity, when just one of the tie carries the whole strength of the node, the disparity approaches 1.
Other measures to quantify the topological diversity in a network have also been used, as the {\it Shannon Entropy} $H_{i}$  or the {\it R\'enyi Disparity} $D_{i}(\gamma)$, where $\gamma$ is a tunable constant (\citealt{Eagle1,Lee}). These quantities however are strongly related to each other; in fact $H_{i}$ behaves like $1/Y_{i}$ and  $D_{i}$ reduces to $1/Y_{i}$ in the case $\gamma=2$, while for $\gamma=1$ it reduces to the Shannon disparity, which is the exponential of the Shannon entropy.

\begin{figure}[t]
\centering
\includegraphics[width=0.6\textwidth]{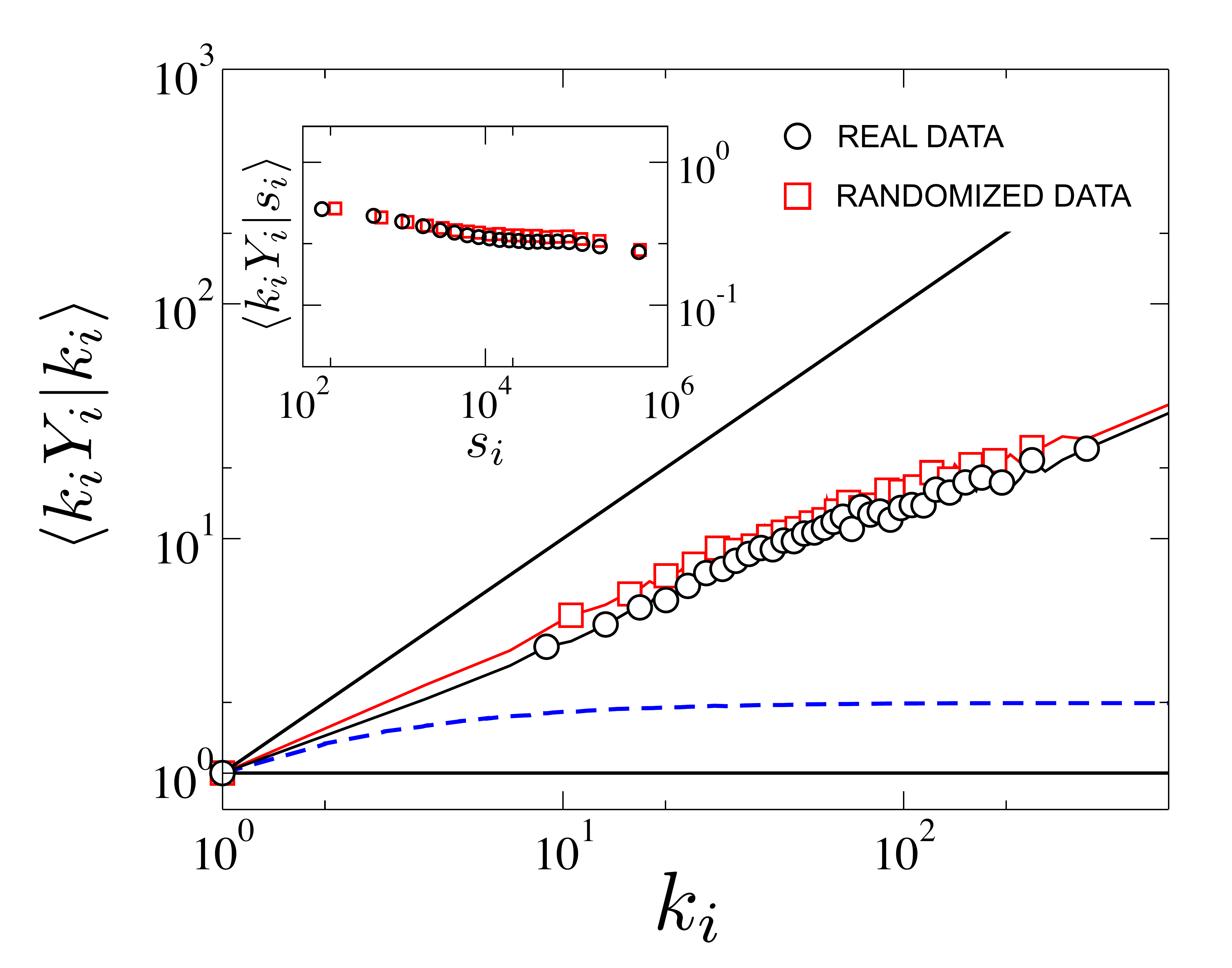}
\caption{(color online) Average $k ~Y_i$ as a function of the social connectivity $k_i$. No significant difference is observed between the real network (black circles) and the randomised one (red squares). Our data are intermediate between the two extreme cases of perfect homogeneity $kY=1$ and perfect heterogeneity $kY=k$ (solid lines), with an exponent of 0.5 for large connectivities. The dashed blue line represents the case in which the weights are distributed according to an exponential distribution, which shows a fast saturation to the homogeneous case. In the inset is shown $k ~Y_i$  as a function of the node's strength $s_i$. No significant correlation is observed between these two quantities, a result obtained also for the randomised network.}
\label{fig3}
\end{figure}

Although the concept of disparity is not a new one (\citealt{Boccaletti,Barthelemy1, Barthelemy2, Lee}), it has received relatively little attention in mobile-phone communication networks. Eagle et al. (2010) examined the diversity of individuals' relationships and their connection with the economic development of communities in which they live. However, the diversity and especially its relationships with its degree and its strength has not been thoroughly investigated for communication networks.
Fig.\ref{fig3} shows $k_{i}Y_{i}$ as a function of the social connectivity $k_{i}$. When all the ties have the same strength, this quantity is $k_{i}Y_{i}=1$ and does not depend on $k_{i}$, while if the distribution is severely heterogeneous we have $k_{i}Y_{i} = k_{i}$.
Our results are intermediate between the two extreme cases of perfect homogeneity and perfect heterogeneity. Specifically, we find that for $k_i>20$ the curve is well fitted by the relation $k_{i}Y_{i} \simeq k_{i}^{\alpha}$ with $\alpha=0.5$.
The exponent $\alpha$ smaller than $1$ indicates a dependence between the social connectivity and the disparity, thus suggesting the existence of different strategies of communication between users who have large numbers of connections and those who have few.

However, as shown in Fig.\ref{fig3}, we find that the same result is obtained after randomising the weights of the ties over the whole network. For a given social connectivity, the disparity of the real case (black circles) is always slightly smaller than the one obtained in its randomised version (red squares), indicating that the real communication is slightly more homogeneous than the one corresponding to a random assignment of tie weights. 
Nevertheless, no significant difference is observed between the real network and the randomised one in the scaling of $Y_{i}$, suggesting that the way users organize their time/attention with each one of his contacts does not alter the  diversity in communication.
This is an important finding since the dependence of the disparity of the flux that passes through a node (in our case the communication time) on its degree or strength is often used to assess the existence of nodes that have different functionalities within the network (\citealt{Almaas, Demontis, Lee}). We rather observe that it is just a reflection of the long tailed nature of the distribution of the weights, which introduces a strong heterogeneity. 
In fact, the larger the connectivity of a node, the higher is the probability that its ties have weights belonging to the tail of the distribution. This is why the observed behaviour differs from the homogeneous  $kY_i=1$ curve.
To test this we calculated the disparity of nodes where the tie weights are now randomly chosen from an exponential distribution with the same mean of the real one.
As shown by the dashed curve in Fig.~\ref{fig3}, in this case we observe a fast saturation of $Y_{i}$ to the homogeneous case in which  $Y_i$ is independent of $k_i$. Note that also in this case, a small deviation from the  $\alpha=1$ behaviour is observed for very small values of $k$. 
Due to this small-size effect we observe an exponent smaller than $1$ in the real data, which is the reason why the curve $k_{i}Y_{i}$  saturates to a constant value for very large values of connectivity. 

A plausible explanation of the observed behaviour of the disparity with $k_i$ comes therefore from the heterogeneity in $w_{ij}$ and the fact that the disparity measure $Y_i$ is very sensitive to the distribution of $w_{ij}$. Actually, if we define ties with small and a large weights respectively as {\it weak} and {\it strong} ties, the result in Fig.~\ref{fig2} indicates that the proportion between strong and weak ties does not change with the social connectivity. Therefore, the larger the social connectivity, the larger the number of strong contacts has to be. Despite the high correlation between the disparity and the social connectivity, the inset of Fig.\ref{fig3} shows that the disparity is mostly independent of the strength of the node. This is due to the fact that the relation between the strength and the social connectivity is not univocal since for a given $s_i$ there are many different values of $k_i$ which in turn give many possible values of disparity.

\section{Conclusions and Discussion}
In this study we examined whether the way in which mobile users distribute their limited time across their social network is related to the size of that network and the intensity of mobile use. There were three key results. First, people with a large personal network spend more time on the phone than people with a small network, which is in line with previous studies (\citealt{Barrat,Goncalves}). However, in correspondence with a threshold value of the size of the contacts network (around 100-150 connections), the total time people can devote to phone communication reaches a maximum. This indicates that a very large number of contacts does not imply a proportional increase in the amount of time invested in communication. Second, we found that the average time people dedicate to their contacts depends on their network size.
In particular, for users with a relatively small number of connections, the time they dedicate to each one of the connections grows proportionally with the network size.
However, there is a decrease in the strength of ties for those users with approximately more than 40 connections. This finding is in line with the Dunbar's theory, which asserts that there is a cognitive limit in the number of social contacts an individual can keep (around 150-200) \cite{Goncalves}.  However, we found that in mobile networks this limit is smaller, probably because beside the cognitive limit, also temporal and  monetary constraints play their role in phone communication. 
Also, given that this analysis was limited to those people with a particular mobile operator, only the members of the personal network who happened to be with that operator would be captured by this analysis, rather than the entire personal network.
Another reason might be the fact that some contacts are seen face-to-face rather than called on mobile phones.
Finally, the limit of 10-40 that we observe may reflect the maximum number of more intense relationships ("affinity layer") which has been found to be lie 30 alters, with the rest above this number ("active layer") being contacted only very occasionally (\citealt{Sutcliffe,Hill2}). 
Moreover, we have seen that this 10-40 persons limit does not depend on the observation period, which indicates that it is a an intrinsic constraint and it is not an effect of the finite time window. Third, we showed that the observed diversity in the way users distribute their time across their network is a reflection of the the long-tailed nature of the distribution of the aggregated time dedicated to each contact. A comparison with a null model indicated that this does not appear to reveal a specific strategy in communication and that the proportion of strong and weak ties does not change with social connectivity. 

Of the three possibilities for how the time spent on mobile phones affects the way in which people distribute their time across their network, these results are most supportive of the 'no change' model. That is, people do not appear to fundamentally alter the way in which they distribute their time across their network in a way that is related to their intensity of mobile use. All users, whether they have high or low intensity of mobile use, distribute their limited time very unevenly across their network, devoting a large part of their time to a small number of contacts. This is in line with a broad range of findings on both human (\citealt{Marlow, Plickert, Roberts3}) and primate sociality (\citealt{Crockford, Wittig}). The current results suggest that whilst mobile phones offer the technical capability  -or affordance- to contact everyone in a user's personal network with equal ease, in fact users still focus the majority of their time on a very limited number of contacts. Thus mobile users do not appear to use the ease of communication to strategically build up their "weak ties"; instead they focus the majority of their communication on a small number of strong ties, as is the case with everyday face-to-face communication (Milardo, Johnson and Huston, 1983), and other modes of communication such as text messaging (\citealt{Reid}) and social network sites (\citealt{Marlow}).
 
This research extends previous work focusing on large mobile phone networks (\citealt{Onnela2}) by examining how the size of an individual's network is related to the intensity of mobile use and how individual users within the network distribute their limited amount of time across their contacts. 
This extends previous works by Roberts and Dunbar (\citealt{Roberts1}), who found a negative relationship between network size and mean emotional closeness. What the current results suggest is that this negative relationship may only apply to those with unusually large networks. In effect, over a certain network size, time constraints may start to limit tie weights, but below this limit mobile users are able to add people to their network without a drop off in tie weights because in this phase their time is unconstrained.
Whilst disparity has been investigated in other types of networks, to the best of our knowledge is has not been examined in a large mobile phone dataset. There are important differences between social networks and other types of networks, and one way in which these networks are different is that there are constraints on social networks (\citealt{Roberts2}) that may not apply to other networks. Thus, with an air transportation network, the volume of airport traffic grows faster than its degree, whereas the converse is true for social networks: the current study and related findings show that the tie weights tends to decrease with increasing network size (\citealt{Roberts1}).

One of the key strengths of this study is that it is based on a very large dataset of 20 million mobile users and 9 billion calls over a very large period of time (almost one year). This type of dataset gives a level of detail, and is on a scale, not possible to achieve with more traditional types of social network studies reliant on questionnaires or interviews (\citealt{Lazer, Watts2}). Thus these results are not reliant on a limited, often student sample (\citealt{Henrich}), and have a high degree of generalizability to at least the European country in which the mobile company operates. However, there were some important limitations to the study. First, in common with other studies reliant on a single mobile operator (\citealt{Onnela2, Palla}), we only included calls in which both the calling and receiving number belonged to the mobile operator under consideration. 
This filtering is needed to eliminate the bias between the operator and other mobile providers since we have full access to the call records of the operator but only partial access to the records of other providers.
As a consequence, the personal network of a user may not reflect his actual personal network as in their complete set of ties to family and friends, (\citealt{Bernard2}), but only the subset of this personal network that is on the same mobile operator as the user. However, the pattern of results, in terms of the uneven distribution of effort across the network and the drop off in tie weights for large networks, is broadly in line with previous findings in this area (\citealt{Marlow, Reid, Roberts3}), and there is no particular reason to suppose that these findings would not be replicated if the calls to the entire personal network could be analysed. Second, we relied on the volume of communication between two individuals as a measure of the weights of the tie. This assumption is well supported in the literature which shows a strong relationship between the frequency or volume of communication between two individuals and the emotional intensity of the tie (\citealt{Hill2, Roberts3, Roberts4, Wellman3,Gilbert}). Studies combining detailed communication records with more subjective judgements of tie weights (\citealt{Eagle1}) offer the possibility of a more comprehensive understanding of how communication patterns are related to tie weights and friendship patterns.

Finally, the direction of the causal relationship between intensity of mobile use and network size was not explored in this study. Does spending lots of time on the phone lead to a larger network, or do people with a pre-existing large network tend to spend more time on the phone? In mobile phone networks, it is reasonable to assume that the topology of the network comes first: as mobile numbers are not typically publicly available, to phone someone you have to acquire their number, which is usually through personal contact. Thus, there is a pre-existing connection between two mobile users before the first call is made.
However, previous longitudinal research has demonstrated that frequent communication is necessary to maintain ties at high levels of emotional intensity (\citealt*{Cummings,Oswald,Roberts3}). Thus over time, changes in the intensity of calls 
between two individuals may well be associated with changes in the emotional intensity of the tie, something that again could be usefully addressed in future studies combining detailed mobile phone data with more subjective judgments about tie weights from the mobile users. 
\section*{Acknowledgments}
We would like to thank Telef\'onica for providing access to the anonymized data. 
E.M. acknowledges funding from Ministerio de Educacion y Ciencia (Spain) through projects i-Math, FIS2006-01485 (MOSAICO), and FIS2010-22047-C05-04. S.R. was supported by an EPSRC  Knowledge Transfer Secondment (KTS) Award; R.D.'s research is supported by a European Research Council Advanced Research Grant. 
\bibliographystyle{model1-num-names}
\section*{References}



\end{document}